\documentstyle[12pt]{article}

\def\Journal#1#2#3#4{{#1}{\bf #2}, #3 (#4)}
\def\NCA{Nuovo Cimento}

\def\NPA{{Nucl. Phys.} A}
\def\PLB{{Phys. Lett.}  B}
\def\PRL{Phys. Rev. Lett.}
\def\PRD{{Phys. Rev.} D}
\def\ZPC{{Zeits.f. Phys.} C}
\def\ZPA{{Zeits.f. Phys.} A}

\def\MPLA{{Mod. Phys. Lett.} A}

\begin{document}

\author{Gary R. Goldstein\footnote{email: ggoldste@tufts.edu} \\
\vspace{-2 mm}
{\small \it Center for Theoretical Physics} \\
\vspace{-2 mm}
{\small \it Laboratory for Nuclear Science} \\
\vspace{-2 mm}
{\small \it and Department of Physics} \\
\vspace{-2 mm}
{\small \it Massachusetts Institute of Technology} \\
\vspace{-2 mm}
{\small \it Cambridge, MA 02139 USA} \\
\vspace{-2 mm}
{\small \it and} \\
\vspace{-2 mm}
{\small \it Department of Physics}\footnote{Permanent address} \\
\vspace{-2 mm}
{\small \it Tufts University} \\
\vspace{-2 mm}
{\small \it Medford, MA 02155 USA}}

\title{Spin Dependent Fragmentation Functions for Heavy Flavor Baryons and
Single Heavy Hyperon Polarization\thanks{This work is supported in
part by funds provided by the U.S. Department of
Energy (D.O.E.) \#DE-FG02-92ER40702 and \#DF-FC02-94ER40818.} } 
\vspace{-0.5 in}
\date{{\small (MIT-CTP: \#3056  \hfill To be published in: {\it
Proceedings of the International Workshop ``Symmetries and spin'',
Praha-SPIN-2000, Czech. J. Phys., supp. Vol. 51 (2001)} 
\hfill 13 November 2000)}}
\vspace{-0.5 in}
\maketitle
\vspace{-0.5 in}
\begin{abstract}
Spin dependent fragmentation functions for heavy flavor quarks to fragment
into heavy baryons are calculated in a quark-diquark model. The production
of intermediate spin 1/2 and 3/2 excited states is explicity included.  
$\Lambda_b$ , $\Lambda_c$ and $\Xi_c$ production rate and polarization at
LEP energies are calculated and, where possible, compared with experiment.
A different approach, also relying on a heavy quark-diquark model, is
proposed for the small momentum transfer inclusive production of polarized
heavy flavor hyperons. The predicted $\Lambda_c$ polarization is 
roughly in agreement with experiment.
\end{abstract}

\newpage

\section{Introduction}     

The fragmentation of quarks into hadrons has been of considerable
theoretical interest as a means of exploring both the perturbative and
soft regime of QCD. The spin dependence of the fragmentation process is
also of interest because of the light it can shed on the related 
longstanding, puzzling experimental results on how the nucleon spin is
shared by its partons. An important theoretical step toward predicting
fragmentation functions was made several years ago -- it was realized
that renormalization group improved QCD perturbation theory, along with
the non-relativistic constituent quark model of the hadrons, could apply  
to the fragmentation of heavy flavor quarks into heavy flavor
mesons~\cite{chang,braaten}. In a series of papers,
Adamov and Goldstein~\cite{adamov1,adamov3} have extended this
model to the fragmentation of heavy flavor quarks into heavy flavor
baryons via the quark-diquark structure of the baryons. (At the same time
a Russian group~\cite{russians} made a similar extension to
baryons. More recently the light cone expansion of the fragmentation
functions was used by the Amsterdam group~\cite{mulders} with an algebraic
model to generate predictions similar to ours.) In the first section 
following, a summary of the recent extension of this model for heavy
baryon fragmentation to include baryon excited states will be
presented. It is through excited state production that {\it
depolarization} of the heavy quark can occur, as will be shown.

Another longstanding puzzle in hadronic scattering has been the sizeable,
nearly energy independent polarization of inclusively produced
hyperons. Early theoretical expectations were that single polarization
effects should be small~\cite{kane}, since (massless) QCD conserves
chirality. Finite mass corrections
allow some polarization, but the magnitude remains
small~\cite{dharma1}. However, it was
expected that in the case of heavy flavor quarks and baryons the mass
effects begin to produce sizeable polarizations~\cite{dharma3}. This has
been born out by recent data on inclusively produced
$\Lambda_c$~\cite{e791}. In the second section
below I will summarize the model calculations~\cite{lambdac} that confront
this data.

\section{Fragmentation functions for heavy baryons}

\begin{figure}[t]
\begin{center}
\includegraphics{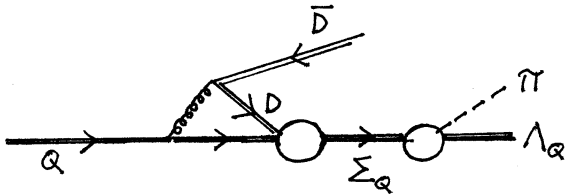}
\end{center}
\vspace{10mm}
\caption{Diagram representing the combined perturbative and soft QCD
calculation of fragmentation through excited baryons.}
\label{fig:diagram}
\end{figure}

The fragmentation of a heavy flavor quark ($Q_1$) into a doubly heavy
flavor meson ($Q_1+\bar{Q}_2$) is in the perturbative regime of QCD;
the required momentum transfer to produce a heavy pair
($Q_2+\bar{Q}_2$) is large compared to the QCD scale $\lambda_{QCD}$. The 
final meson is formed via the non-relativistic potential between the
comoving quark and antiquark. The corresponding picture for baryon
production has the heavy quark pair replaced by a diquark pair, as shown
in Fig.~\ref{fig:diagram}.
The heavy quark emits a hard gluon that creates the diquark pair.
Diquarks can be either scalar or vector, with the vector diquark having
the larger mass (as suggested by the nucleon-$\Delta $ mass difference).
If the diquark is scalar, the heavy quark forms a ground state spin $\frac
12$ baryon with it.
Since the scalar diquark is spinless and the ground state has no orbital
angular momentum, the heavy baryon has nearly the same
helicity as the heavy quark. Vector diquarks have spin 1 and the resulting
helicity of a baryon (composed of a heavy quark and a vector diquark) is
not the same as that of the single heavy quark,
but naturally depends on the helicity of the diquark. That does not
produce any depolarization of the heavy quark by itself - the heavy quark
inside of the baryon does not change its helicity. However, depolarization 
may occur during the ``relaxation'' of the baryon into the ground state.
The depolarizing features of the spin excited baryon decay into the ground
state were discussed by Falk and Peskin \cite{peskin}. Two parameters are
crucial in determining the amount of depolarization relaxation,
$\Delta M$, the mass splitting between the $\frac 32$ and $\frac
12^{\prime}$ excited states of the heavy baryon (which is
related to the spin-dependent part of the QCD interaction with
scale determined by $\frac{\Lambda _{QCD}^2}{m_Q}$ and is responsible for
flipping the quark spin) and the excited baryon lifetimes , 
$\frac 1\Gamma $, which will be nearly the same.
The longer this lifetime is, in comparison to the time required for heavy
quark spin flipping, the more likely it is for the diquark and quark to
mix together forming a randomized spin state. For the heavy mass limit  
$\Delta M$ disappears
and there is no depolarization. For finite mass depolarization occurs.
Recent data on $\Lambda_b$ confirm this
expectation; OPAL~\cite{newlep} finds an average longitudinal polarization
of $-0.56\pm 0.22$ where the Standard Model prediction {\it without}
excited state contributions would be $-0.94$. This and related results
provide the impetus for studying the fragmentation via intermediate
excited states.

The fragmentation functions (at leading twist, i.e. large $Q^2$) that I
will discuss can be related to simple probability functions for quarks of
fixed or mixed helicities to produce ground state (spin $\frac 12$)
baryons with the same or different helicities. Symbolically,
\begin{eqnarray}
\hat{f}_1 & \sim & \left||\lambda_Q = +{\frac 12}> \rightarrow
|\lambda_{B_Q} =
+{\frac 12}>\right|^2 + \left||\lambda_Q = +{\frac 12}> \rightarrow  
|\lambda_{B_Q} =
-{\frac 12}>\right|^2 \nonumber \\
\hat{g}_1 & \sim & \left||\lambda_Q = +{\frac 12}> \rightarrow
|\lambda_{B_Q} =
+{\frac 12}>\right|^2 - \left||\lambda_Q = +{\frac 12}> \rightarrow
|\lambda_{B_Q} =
-{\frac 12}>\right|^2  \nonumber \\
\hat{h}_1 & \sim & \left|||S_T^Q = +{\frac 12}> \rightarrow |S_T^{B_Q} =
+{\frac 12}>\right|^2 - \left||S_T^Q = +{\frac 12}> \rightarrow
|S_T^{B_Q} =
-{\frac 12}>\right|^2 
\label{eq:frags}
\end{eqnarray}
where ${S_T}^Q \sim |+> \pm |->$ is the {\it transversity}, a variable
introduced originally by Moravcsik and Goldstein~\cite{transversity} to
reveal an underlying simplicity in nucleon--nucleon spin dependent
scattering amplitudes. The chiral odd fragmentation function $\hat{h}_1$ 
is of particular interest, since the analogous structure function can
not be measured in deep inelastic scattering.

The perturbative calculation of the baryon $B$ fragmentation functions
proceeds from the partial
width~\cite{braaten} for the inclusive decay process $Z^0\rightarrow B+X$. 
The expression for that width will be simplified if we restrict
ourselves to the fragmentation channel shown in
Fig.~\ref{fig:diagram}. Then, while $z$ is kept fixed, the limit of
large mass of the $Z^0$ along with large energy of the heavy quark, $q_0$,
and the baryon, $l_0$, yields
\begin{equation}
\lim_{l_0\longrightarrow \infty }d\Gamma (Z^0\rightarrow
B(E)+X)=\lim_{q_0\longrightarrow \infty }\int_0^1\!dz\,d\hat{\Gamma}%
\!(Z^0\rightarrow Q(E/z)+X,\mu )\,f_1(z,\mu ).  \label{simple}
\end{equation}
for the spin averaged case, with
\begin{equation}
\int_0^1\!dz\,f_1(z,\mu )=\frac{\Gamma _1}{\Gamma _0},  \label{f1}
\end{equation}
where $\Gamma _1$ is the decay width of $Z^0$ into the ground state baryon
and appropriate remnants - antiquark, spectator diquark and pion, while
$\Gamma_0$ is the total decay width of the $Z^0$ into the heavy quark
pair~\cite{adamov1}.

For the fragmentation into an excited baryon, followed by $\pi$ (or
$\gamma$) decay into the ground state the full $Z^0$ width has multiple
phase space integrations.
\begin{equation}
\Gamma _1=\frac 1{2M_Z}\int
[d\overline{q}][dl][dp^{\prime}][d\overline{\pi
}](2\pi )^4\delta ^4(Z-\overline{q}-l-p^{\prime }-\overline{\pi })\left|
M_1\right| ^2
\label{gam1}
\end{equation}
where $\bar{q}$, $l$, $p^{\prime }$ and $\pi $ are the 4-momenta of the 
$\bar{Q}$, $B_Q$,$\overline{D}$ and the pion (or photon), respectively,
and the amplitude
$M_1$ is summed and averaged over unobserved spins and colors. 

In order to factor the fictitious decay width $\Gamma_0$ out of
Eq.~\ref{gam1} we transform the phase space variables to
production independent variables $x_1=\frac{p_0+p_L}{q_0+q_L}$ and
$x_2=\frac{l_0+l_L}{p_0+p_L}$ that can be loosely thought of as Feynman
scaling
variables for each subprocess, i.e. excited baryon production and
decay. We also introduce a further simplification - the ratio of the
narrow decay width to the mass of the excited baryons allows the narrow
width approximation to be used. 
The resulting phase space integral can be written as:
\begin{eqnarray}
\Gamma _1 & = & \frac 1{2M_Z}\frac 1{256\pi ^4}\int
[d\overline{q}][dq](2\pi  
)^4\delta ^4(Z-q-\overline{q})  \nonumber \\
& & \cdot \int ds_q\theta \left( s_q-\frac{M_\Sigma
^2}z-\frac{m_d^2}{1-z}%
\right) \int d\phi d\varphi dx_1dx_2\left| A_1\right| ^2
\end{eqnarray}
Here $\left| A_1\right| ^2\delta (p^2-M^2)=\left| M_1\right| ^2$. The two
angles $\phi$ and $\varphi$  are
associated with the position of the transverse momentum
vector in two frames of reference. The first is the frame determined by
the three-momentum of the heavy quark and a fixed vector perpendicular to
it, which is arbitrary unless it is the heavy quark's transverse spin
vector (which enters in $h_1$ only). The angle $\phi$ is the azimuthal
angle between this plane and the transverse momentum vector of the excited
baryon. The second plane is constructed out of the
three-momentum of the excited baryon and the spin vector perpendicular to
that three-momentum but having no transverse component relative to the
first frame. The second angle $\varphi$ is defined as the azimuthal angle
between this latter plane and the transverse momentum  of the baryon.

After factoring out the production decay width we are left with the
somewhat simpler expression for $f_1$:
\begin{equation}
f_1(z,\mu)={\frac 1{16\pi ^2}}\lim_{q_0\rightarrow \infty
}\int_{s_{th}}^\infty \!ds\frac{dx_2}{x_2}{\frac{\left| A_1\right|^2}{%
\left| M_0\right|^2}}
\label{eq:ratio}
\end{equation}
The spin dependent fragmentation functions can be obtained
directly, using the modified Eq.\ref{eq:ratio}:
\begin{eqnarray}
g_1(Q,z) &=&{\frac 1{256\pi ^4}}\lim_{q_0\rightarrow \infty
}\int_{s_{th}}^\infty \!ds\frac{dx_2}{x_2}d\phi d\varphi {\frac{\left|
A_{1+}\right| ^2-\left| A_{1-}\right| ^2}{\left| M_0\right| ^2}}
\label{eq:spin} \\
h_1(Q,z) &=&{\frac 1{256\pi ^4}}\lim_{q_0\rightarrow \infty
}\int_{s_{th}}^\infty \!ds\frac{dx_2}{x_2}d\phi d\varphi {\frac{\left|
A_{1y+}\right| ^2-\left| A_{1y-}\right| ^2}{\left| M_0\right| ^2}}
\label{eq:spinh1}
\end{eqnarray}
with new indices specifying the spin alignment ($|y+> \sim |+>+i|->$).
Angular integration is
especially complicated for $h_1$, because matrix elements involve spin
projections that make them no longer azimuthally symmetric.

The expressions above are general for the four body final state. The model
dependence is hidden inside of the $\left| A_1\right|^2$ with the delta
function obtained in the narrow width approximation integrated out. The
final ground state baryon can be produced via one of the two intermediate
states.  The amplitudes for both $\frac 12$ and $\frac 32$ have been
obtained (details can be found in \cite{adamov3}).

The amplitudes for $\frac 12$ and $\frac 32$ depend on the
gluon--vector
diquark pair production amplitude, which involves chromodynamic diquark
currents~\cite{gold1} and form factors. The important contributions
yield the chromoelectric part of the matrix element contributing
to the spin ${\frac {1}{2}}^\prime$ baryon
\begin{equation}
A_{E\,1/2}=-\frac{\psi(0)}{\sqrt{3m_d}}F_E(k^2)
\gamma_5\gamma^{\mu}
\frac{1+\mathbf{v}}{2}g_s\bar{\epsilon}^{*}_{\mu}[k_{\lambda}-
2m_d v_{\lambda}]P^{\lambda}.
\label{eq:E1/2}
\end{equation}
and
\begin{equation}
A_{E\,3/2}^{\nu}=-\frac{\psi(0)}{\sqrt{2m_d}}F_E(k^2)
g_s\bar{\epsilon}^{*\nu}[k_{\lambda}-
2m_d v_{\lambda}]P^{\lambda},
\label{eq:E3/2}
\end{equation}
where
\begin{equation}
P^{\lambda}=
\bigtriangleup^{\lambda \nu}g_s\gamma_{\nu}
\frac{m_Q(1+\mathbf{v})+\mathbf{k}}{(s-m^2_Q)}\Gamma.
\label{eq:Plambda}
\end{equation}
for the spin $\frac{3}{2}$. The wavefunctions for the diquark--quark
baryon formation are obtained from the power law potential of
Eichten and Quigg~\cite{quigg}.

The remaining calculations of the amplitudes are straightforward, but
quite complex. The squares of the matrix elements involve the trace of various
amplitudes and spin projection operators ($\frac{1+\gamma _5\not{S}_\alpha
}2)$ with $\alpha $ being the spin projection index and $S_\alpha $ being
the spin four-vector corresponding to that projection. The two cases for
this problem are the longitudinal or helicity and transverse spin
vectors. After simplifying the squares of matrix elements and leaving only
the leading terms in the high momentum limit, scalar products of all the  
involved four-momenta and spin vectors remain. The angular dependences can
then be integrated over. 

The integrations produce spin-dependent fragmentation functions that are
defined at the
scale $\mu_0 = m_Q + m_{diquark}$. To evolve them to higher values of the
defining scale (or the typical $Q^2$) we utilize the appropriate
spin-dependent Altarelli-Parisi integro-differential equations as
determined by Artru and Mekhfi~\cite{artru}.

\begin{figure}[t]
\begin{center}
\includegraphics{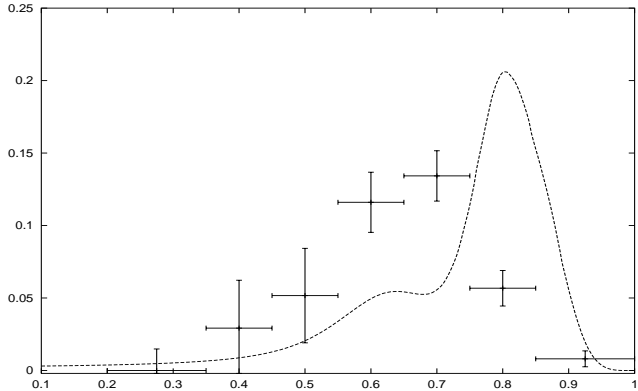}
\end{center}
\vspace{40mm}
\caption{Spin independent function $f_1(z)$ for $\Lambda_c$ evolved to
$\mu$= 5.5 GeV. The data are from CLEO~\cite{cleo1}.}
\label{fig:c_data}
\end{figure}

\begin{figure}[t]
\begin{center}
\includegraphics{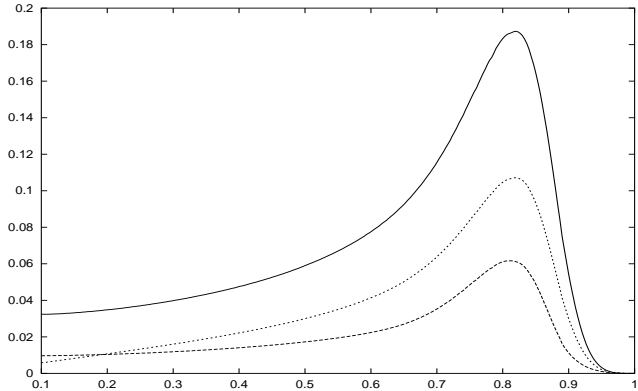}
\end{center}
\vspace{40mm}
\caption{Fragmentation functions for $\Lambda_b$ evolved to 45 GeV. At the
peak $f_1$ is largest, followed by $h_1$ and $g_1$.}
\label{fig:b_ev}
\end{figure}

The resulting $\Lambda_c$ $f_1(z)$ evolved to 5.5 GeV is shown in
Fig.~\ref{fig:c_data}. The data are probably at the low edge of reliable
energy for our evolution. Predictions for 45 GeV are in
ref.~\cite{adamov3}. The corresponding $\Lambda_b$ is given in
Fig.~\ref{fig:b_ev}. To compare with yield measurements at LEP $f_1$ is
integrated over $z$. The results are in Table~\ref{tb:tot} and agree with
data.
 
\begin{table}[b]
\caption{Total Fragmentation Probabilities}
\vspace{-1mm}
\small
\begin{center}
\begin{tabular}{|c|c|c|}
\hline
\raisebox{0mm}[4mm][2mm]{Particle} & Experiment & Prediction \\
\hline\hline
$\Lambda_c$ & 5.6$\pm$ 2.6\% [OPAL~\cite{opal}]& 3.9\%  \\
$\Xi_c$     &                      & 0.59\% \\
$\Lambda_b$ & 7.6$\pm$ 4.2\% [ALEPH~\cite{aleph}]&6.7\% \\ 
\hline
\end{tabular}
\vspace{-1mm}
\end{center}
\label{tb:tot}
\end{table}

\section{Polarization of inclusively produced heavy flavor baryons}

Several years ago, with W.G.D. Dharmaratna, I developed a hybrid model for
hyperon polarization in inclusive reactions~\cite{dharma1}. The model
involved the order $\alpha_s^2$ QCD perturbative calculation of strange
quark polarization due to the hard QCD subprocesses. The
interference between tree level and one loop diagrams gives rise to
significant, albeit small polarization. Of
particular importance for describing then existing data on $p+p\rightarrow
\Lambda + X$ were one loop diagrams leading to strange quark pair
production from gluons or quark-antiquark pairs, with the 
gluon fusion being more significant. For strange quarks, their low current
or constituent quark mass (compared to $\Lambda_{QCD}$) made the
application of PQCD marginal.  Nevertheless, it was
realized that the hadronization process, by which the polarized strange
quark would recombine with a (ud) diquark system to form a $\Lambda$, was
crucial for understanding the subsequent hadron polarization. A simple  
prescription was introduced to ``pull'' or accelerate the negatively
polarized, relatively  slow s-quark along with a fast moving diquark
resulting from a pp collision to form the hadron with particular $x_F$ and
$p_T$. This recombination prescription is similar to the ``Thomas
precession''model~\cite{degrand},
which posits that the s-quark needs to be accelerated by a confining
potential or via a ``flux tube''~\cite{lund} at an angle 
to its initial momentum in order to join with the diquark to form the
hyperon. The skewed acceleration gives rise to a spin precession for the
s-quark. That latter enhances the s-quark's polarization by the Thomas
precession. The hybrid model combines hard perturbative QCD with a simple
model for non-perturbative recombination.
It accounted for the contemporaneous polarization data which were
determined for a range of $x_F$ and $p_T$ values and the predicted
kinematic dependence was confirmed, along with the nearly negligible
overall energy dependence of the polarization. However, the PQCD based  
hybrid model  
is best tested in the production of heavy flavor hadrons, wherein the
heavy quark needs to be produced at large energies compared to the
$\Lambda_{QCD}$ scale. The polarization in the QCD subprocesses was
calculated~\cite{dharma3} for all flavors and  indeed, the polarization  
increases substantially with constituent mass; the peak polarization goes
roughly as the mass. 

\begin{figure}[t]
\begin{center}
\includegraphics{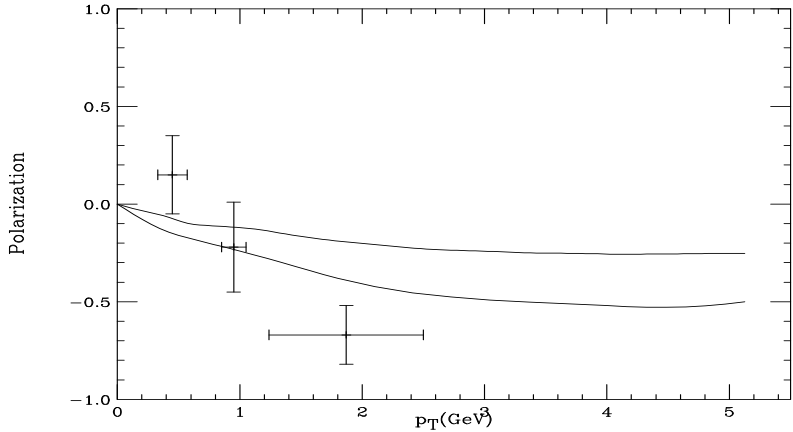}
\end{center}
\vspace{48mm}
\caption{$Lambda_c$ polarization from $\pi^{-} p\rightarrow
\Lambda_c + X$. The larger polarization includes heavy mass enhancements.
The data~\cite{e791} are from E791.}
\label{fig:pol}
\end{figure}

I applied this reasoning to $\pi + p \rightarrow \Lambda_c
+ X$. It is known that $\pi$ induced inclusive $\Lambda$ production also
produces significant polarization of $-28\pm10\%$ for $p_T$ values near
and  above 1~GeV/c~\cite{accmor}. In this process the ud-diquark system
that combines with the strange or heavy flavor polarized quark must be
pulled from the target proton or the pion sea. Various possibilities were
considered in determining the
polarization for $\Lambda_c$ via $\pi$ production, where new data
exist~\cite{e791}. I adopted the same recombination
prescription used for the $p+p \rightarrow \Lambda + X$.

The basic equation for production is given by
\begin{equation}
d^2\sigma(\uparrow {\mathrm and} \downarrow)/dx_Qdp_T =
\sum_{i,j}\int_0^1 dx_1\int_0^1 dx_2 f_i^{p,\pi}(x_1)f_j^p(x_2)
d^2\sigma(\uparrow {\mathrm and} \downarrow)/dx_Qdp_T.
\label{dsigma}
\end{equation}
Next the algebraic recombination formula is applied to obtain
the corresponding $\Lambda_Q$ polarized cross section at $x_F(=a+bx_Q)$
and $p_T$. 

The full dependence on both variables $x_F, p_T$ is similar to the
$\Lambda$ data. The overall polarization of $\Lambda_c$ in
fig.~\ref{fig:pol} is consistent with the new data. It will be of  
considerable interest for the hybrid model to see how well this detailed
behavior will be confirmed when more data are available.

\bigskip
{\small The author is grateful for the invitation to SPIN 2000 and
especially to the organizers.}

\end{document}